\documentclass[prl,floatfix,superscriptaddress,twocolumn,showpacs]{revtex4-1}

\usepackage{graphicx}
\usepackage{dcolumn}
\usepackage{bm}
\usepackage{lipsum}

\newcommand{\mv}[1]{\ensuremath{\mathbf{#1}}} 
\newcommand{\gv}[1]{\ensuremath{\mbox{\boldmath$ #1 $}}} 
\newcommand{\avg}[1]{\left \langle #1 \right \rangle} 

\usepackage[colorlinks]{hyperref}  
\usepackage{cleveref}
\hypersetup{linkcolor=blue,citecolor=blue,urlcolor=blue}

\usepackage{titlesec}
\titleformat{\section}
  {\centering\normalfont\fontsize{11}{15}\bfseries}{\thesection}{1em}{}

\usepackage[toc]{appendix}

\crefname{figure}{fig.}{figs.}
\Crefname{figure}{Figures}{Figures}

\begin{document}

\setcounter{page}{1} 

\title{Cage Length Controls the Non-Monotonic Dynamics of Active Glassy Matter}

\author{Vincent E. Debets}
\author{Xander M. de Wit}
\author{Liesbeth M.C. Janssen}
\email{l.m.c.janssen@tue.nl}
\affiliation{Department of Applied Physics, Eindhoven University of Technology, P.O. Box 513,
5600 MB Eindhoven, The Netherlands}
\affiliation{Institute for Complex
Molecular Systems, Eindhoven University of Technology, P.O. Box 513, 5600 MB Eindhoven, The Netherlands\\}

\begin{abstract}%
{\noindent Dense active matter is gaining widespread interest due to its remarkable similarity with conventional glass-forming materials. However, active matter is inherently out-of-equilibrium and even simple  models such as active Brownian particles (ABPs) and active Ornstein-Uhlenbeck particles (AOUPs) behave markedly differently from their passive counterparts. Controversially, this difference has been shown to manifest itself via either a speedup, slowdown, or non-monotonic change of the glassy relaxation dynamics. Here we rationalize these seemingly contrasting views on the departure from equilibrium by identifying the ratio of the short-time length scale to the cage length, i.e.\ the length scale of local particle caging, as a vital and unifying control parameter for active glassy matter. In particular, we explore the glassy dynamics of both thermal and athermal ABPs and AOUPs upon increasing the persistence time. We find that for all studied systems there is an optimum of the dynamics; this optimum occurs when the cage length coincides with the corresponding short-time length scale of the system, which is either the persistence length for athermal systems or a combination of the persistence length and a  diffusive length scale for thermal systems. This new insight, for which we also provide a simple physical argument, allows us to reconcile and explain the manifestly disparate departures from equilibrium reported in many previous studies of dense active materials. 
}
\end{abstract}

%


\maketitle 
\noindent \textit{Introduction --} Throughout the previous decade, active matter has attracted increasingly wide attention in the field of soft matter and biophysics \cite{Bechinger2016,Ramaswamy2010,Marchetti2013rev}. Consisting of particles that continuously convert energy into mechanical work or autonomous motion, active materials are naturally outside the realm of thermodynamic equilibrium and as a result exhibit a plethora of surprising features.
Although initial active matter studies have spanned mostly from the single-particle level up to the moderate-density regime \cite{Bechinger2016}, more recent interest has shifted towards high-density active matter systems \cite{Janssen2019active,Berthier2019review}. This has already yielded a significant amount of work on monodisperse systems in the context of crystallization and phase separation~\cite{Briand2016,Briand2018,Geyer2019,Digregorio2018,Caporusso2020}. Another important catalyst for this refocused interest is the striking similarity between dense disordered active matter and conventional (passive) glassy materials \cite{Debenedetti2001supercooled,Janssen2019active,binder2011}. Indeed, glassy phenomenology in active matter has been observed in, e.g., experiments of synthetic colloids \cite{Klongvessa2019colloid1,Klongvessa2019colloid2} and living cells \cite{Angelini2011cell,Garcia2015cell,Nishizawa2017cell,Parry2014bacterial,Zhou2009cell}, as well as in simulation and theory \cite{Voigtmann2017active,SzamelABP2019,SzamelAOUP2015,SzamelAOUP2016,FengHou2017,BerthierABP2014,DijkstraABP2013,BerthierAOUP2017,Berthier2013activeglass,Flenner2020,FlennerAOUP2016,Henkes2011active,Reichert2020modecoupling,Reichert2020tracer,Reichert2021rev,Nandi2018,Sollich2020,janzen2021aging,Janssen2017aging,Bi2016cell,Velasco2020}. Although dense systems are dominated by interactions and could in principle be expected to be fairly similar across passive and active matter, activity still plays a non-trivial role~\cite{BerthierAOUP2017,SzamelAOUP2015,Flenner2020,FlennerAOUP2016,DijkstraABP2013,Berthier2013activeglass}.
What this role precisely is and to what degree active systems can be mapped onto passive ones has therefore emerged as an important new area of research.

Two simple model systems, which are widely used in theoretical and simulation studies of dense active matter, are so-called active Brownian particles (ABPs) \cite{Romanczuk2012active,Ramaswamy2017active,Lowen2020active,SzamelABP2019,Marchetti2012active,Hagen2011ABP} and active Ornstein-Uhlenbeck particles (AOUPs) \cite{Szamel2014AOUP,Maggi2015AOUP,Fodor2016AOUP}. These models differ in the manner in which they model active forces, either describing them as forces with a constant magnitude undergoing rotational diffusion (ABPs), or letting them evolve in time via an Ornstein-Uhlenbeck process (AOUPs). However, even for these relatively simple model active particles, the departure from equilibrium in dense systems is confounded by surprising and seemingly contrasting results.
Notably, in several studies the long-time particle dynamics has been shown to change non-monotonically upon increasing the persistence of the constituent particles \cite{FlennerAOUP2016,SzamelAOUP2015,BerthierABP2014,BerthierAOUP2017}, while other works find either monotonically enhanced \cite{DijkstraABP2013,Voigtmann2017active,BerthierAOUP2017,Mandal2016} or decreased \cite{BerthierAOUP2017,Flenner2020} dynamics. To account for the change in dynamics, it was recently proposed by Liluashvili \textit{et al.}\  \cite{Voigtmann2017active} that the so-called cage length $l_{\mathrm{c}}$ \cite{Kob2002supercooled,Janssen2018front}, i.e.\ the space each particle is permitted before encountering its neighboring particles, might be a crucial length scale 
that provides an offset
beyond which active motion influences glassy behavior in thermal hard-sphere systems.

Here we show that the cage length is an even more important parameter than previously suggested, and in fact holds the key to rationalizing and reconciling the apparently disparate views on the departure from thermal equilibrium for both thermal and athermal dense active systems.
Briefly, we explore the dynamics of interacting ABPs and AOUPs upon increasing the persistence time, while at the same time fixing their effective temperature. For all considered settings, we retrieve a non-monotonic dependence of the long-time diffusion coefficient whose qualitative shape, consisting of an initial increase and later decrease, also remains the same. By replacing the persistence time by the ratio of the relevant short-time length scale (either the sole persistence length for athermal systems or a combination of the persistence length and a diffusive length scale for thermal systems) to the cage length as our control parameter, we find the optimum of the dynamics in all cases to coincide with a value equal to one. We discuss how this can explain a large number of previous findings and thus establish this ratio as the central and unifying dimensionless length scale for active glassy materials.


\noindent \textit{Simulation Details --} As our model system we take a three-dimensional (3D) Kob-Andersen binary mixture consisting of $N_{\mathrm{A}}=800$ and $N_{\mathrm{B}}=200$ quasi-hard self-propelling spheres of type A and B respectively. Each particle $i$ is described by the following overdamped equation of motion \cite{Farage2015,DijkstraABP2013,FengHou2017}
\begin{equation}
    \dot{\mv{r}}_{i} = \zeta^{-1} \left( \mv{F}_{i} + \mv{f}_{i} \right) + \gv{\xi}_{i},
\end{equation}
where $\mv{r}_{i}$ denotes the position of particle $i$, $\zeta$ the friction coefficient, $\mv{F}_{i}$ and $\mv{f}_{i}$ the interaction and self-propulsion force acting on particle $i$ respectively, and $\gv{\xi}_{i}$ a Gaussian noise with zero mean and variance $\avg{\gv{\xi}_{i}(t)\gv{\xi}_{j}(t^{\prime})}_{\mathrm{noise}}=2k_{B}T\zeta^{-1}\mv{I}\delta_{ij}\delta(t-t^{\prime})$, with $k_{B}T\equiv T$ the thermal energy (temperature), $t$ the time, and $\mv{I}$ the unit matrix. The interaction force $\mv{F}_{i}=-\sum_{j \neq i} \nabla_{i} V_{\alpha\beta}(r_{ij})$ is derived from a quasi-hard-sphere powerlaw potential $V_{\alpha\beta}(r)= 4\epsilon_{\alpha\beta}\left( \frac{\sigma_{\alpha\beta}}{r}\right)^{36}$ \cite{Weysser2010structural,Lange2009} and the interaction parameters, i.e.\ $\epsilon_{\mathrm{AA}}=1,\  \epsilon_{\mathrm{AB}}=1.5,\  \epsilon_{\mathrm{BB}}=0.5,\  \sigma_{\mathrm{AA}}=1,\ \sigma_{\mathrm{AB}}=0.8,\  \sigma_{\mathrm{BB}}=0.88$, are, in combination with setting the friction coefficient to unity $\zeta=1$, chosen to give good glass-forming mixtures \cite{Kob1994,Michele2004}.

The distinction between ABP and AOUP models rests in their time evolution of the self-propulsion force $\mv{f}_{i}$. For ABPs the absolute value of the force $f$ remains constant in time, i.e.\ $\mv{f}_{i}=f\mv{e}_{i}$, while the orientation $\mv{e}_{i}$ undergoes rotational diffusion \cite{Farage2015,FengHou2017},
\begin{equation}
    \dot{\mv{e}}_{i} = \gv{\chi}_{i} \times \mv{e}_{i},
\end{equation}
subject to a Gaussian noise process with zero mean and  variance $\avg{\gv{\chi}_{i}(t)\gv{\chi}_{j}(t^{\prime})}_{\mathrm{noise}}=2D_{\mathrm{r}}\mv{I}\delta_{ij}\delta(t-t^{\prime})$ whose amplitude is determined by the rotational diffusion coefficient $D_{\mathrm{r}}$.
In comparison, for AOUPs the self-propulsion force evolves in time according to \cite{SzamelAOUP2016,FlennerAOUP2016,SzamelAOUP2015,BerthierAOUP2017,Flenner2020,FengHou2017}
\begin{equation}
    \dot{\mv{f}}_{i} = \tau^{-1} \mv{f}_{i} + \gv{\eta}_{i},
\end{equation}
Here, $\tau$ depicts the characteristic decay time of the self-propulsion and $\gv{\eta}_{i}$ is an internal Gaussian noise process with zero mean and variance  $\avg{\gv{\eta}_{i}(t)\gv{\eta}_{j}(t^{\prime})}_{\mathrm{noise}}=2D_{f}\mv{I}\delta_{ij}\delta(t-t^{\prime})$ governed by a diffusion coefficient $D_{f}$.

If we neglect particle interactions, both models yield a persistent random walk (PRW) with mean square displacement (MSD) \cite{FengHou2017}
\begin{equation}\label{MSDsingle}
    \avg{\delta r^{2}(t)}=6Tt + 6T_{\mathrm{a}} \left(\tau_{\mathrm{p}}(e^{-t/\tau_{\mathrm{p}}} - 1) + t \right).
\end{equation}
Such a PRW is characterized by a persistence time $\tau_{\mathrm{p}}=(2D_{\mathrm{r}})^{-1}$ (ABP), $\tau_{\mathrm{p}}=\tau$ (AOUP), an active temperature $T_{\mathrm{a}}=f^{2}\tau_{\mathrm{p}}/3$ (ABP), $T_{\mathrm{a}}=D_{f}\tau^{2}_{\mathrm{p}}$ (AOUP), and the (passive) temperature $T$. In particular, at short times ($t\ll \tau_{\mathrm{p}}$) the motion is comprised of a diffusive and ballistic contribution $\avg{\delta r^{2}(t)}\approx 6Tt + 3T_{\mathrm{a}}t^{2}/\tau_{\mathrm{p}}$, and
in the long-time limit ($t\gg \tau_{\mathrm{p}}$) it becomes fully diffusive with an enhanced diffusion coefficient $\avg{\delta r^{2}(t)}\approx 6(T_{\mathrm{a}}+T)t\equiv 6T_{\mathrm{eff}}t$. 
Moreover, in the limit $\tau_{\mathrm{p}}\rightarrow 0$ (with $T_{\mathrm{a}}\sim \mathrm{constant}$), both models become equivalent to a Brownian system at a temperature equal to the effective temperature $T_{\mathrm{eff}}=T_{\mathrm{a}}+T$. 
To compare both models we will take as our control parameters $T$, $\tau_{p}$, $T_{\mathrm{eff}}$, and the number density $\rho$.

Simulations are carried out using LAMMPS \cite{Lammps}. We impose periodic boundary conditions, fix the cubic box size to set the number density, let the system run sufficiently long to ensure that no significant aging takes place, and afterwards track the particles over time. All results are presented in reduced units where $\sigma_{\mathrm{AA}}$, $\epsilon_{\mathrm{AA}}$, $\epsilon_{\mathrm{AA}}/k_{\mathrm{B}}$, and $\zeta\sigma^{2}_{\mathrm{AA}}/\epsilon_{\mathrm{AA}}$ represent the units of length, energy, temperature, and time respectively \cite{Flenner2005}. For more details on the simulation protocol we refer to the Supplementary Information (SI).

\noindent \textit{Athermal Active Particles --} For simplicity, we initially focus on athermal systems ($T=0$, $T_{\mathrm{eff}}=T_{\mathrm{a}}$), 
choose three state points $[T_{\mathrm{eff}}=2.448,\ \rho=1.25]$, $[T_{\mathrm{eff}}=1.5,\ \rho=1.2]$, $[T_{\mathrm{eff}}=0.528,\ \rho=1.1]$ where the systems exhibit mildly supercooled behavior, and vary the persistence time $\tau_{\mathrm{p}}$ to study the departure from equilibrium. An additional advantage of the chosen state points resides in the self-similar nature of the powerlaw potential. This implies that for such a potential the behavior of a passive Brownian system is fully characterized by the parameter $\Gamma=T\rho^{-12}$ \cite{Weysser2010structural,Michele2004,Hansen2013simple}. More concretely, since $\Gamma$ (using $T_{\mathrm{eff}}$ instead of $T$) is the same for all three studied state points, they should yield equivalent dynamics when we take the limit $\tau_{\mathrm{p}}\rightarrow 0$, allowing for a convenient comparison. 

\begin{figure}[ht!]
    \centering
    \includegraphics [width=9cm,height=3.9cm] {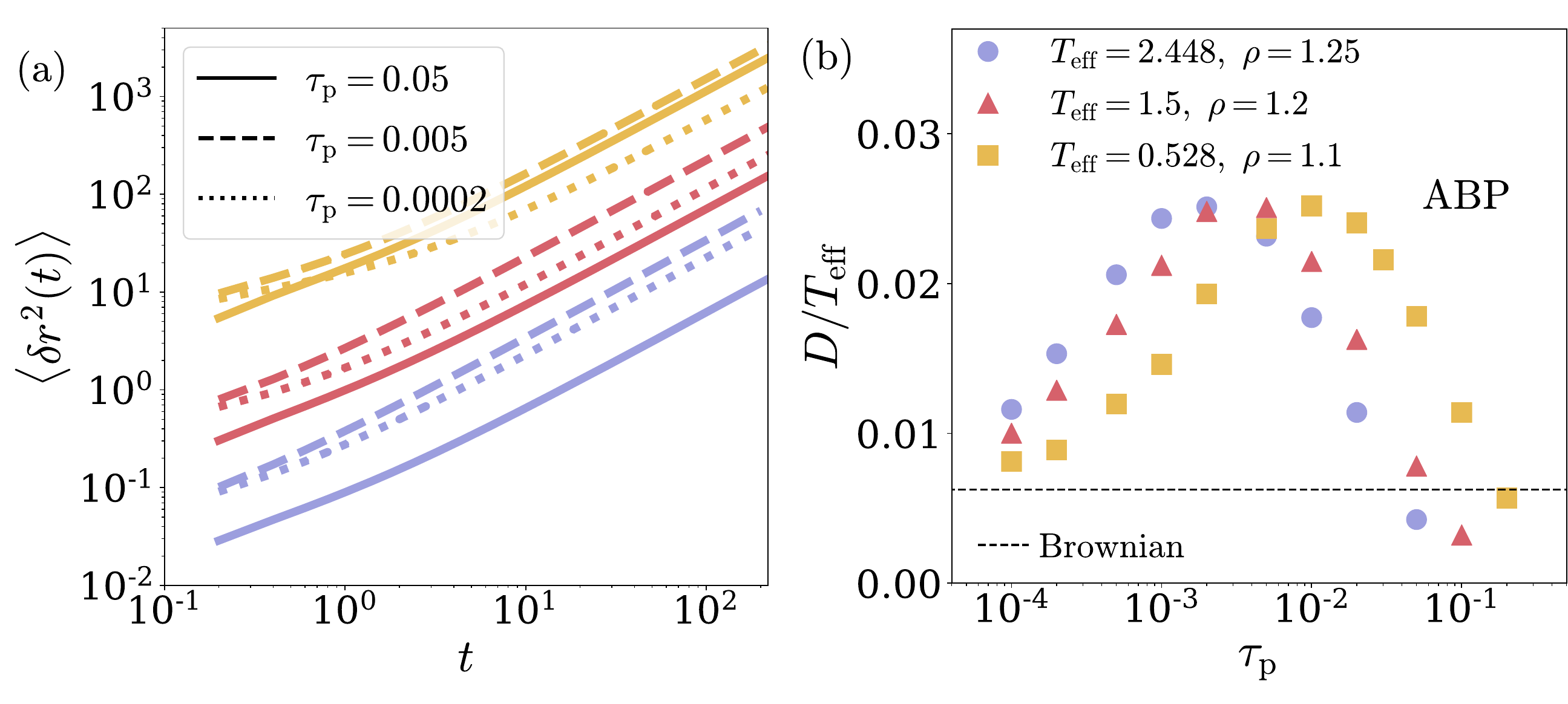} 
    \caption{(a) MSDs of athermal ABPs for different persistence times $\tau_{\mathrm{p}}$. The results for $[T_{\mathrm{eff}}=2.448,\ \rho=1.25]$ (blue), $[T_{\mathrm{eff}}=1.5,\ \rho=1.2]$ (red), and $[T_{\mathrm{eff}}=0.528,\ \rho=1.2]$ (orange) are multiplied by a factor $1$, $10$, and $200$ respectively for visibility.
    (b) The corresponding self-diffusion coefficients normalized by the effective temperature. 
    Upon increasing $\tau_{\mathrm{p}}$ the dynamics initially speeds up before slowing down and dropping below the value obtained from equivalent ($T=T_{\mathrm{eff}}$) passive Brownian dynamics simulations (dashed line).  }
    \label{Fig1}
\end{figure}

Starting with the ABP model, we study its $T=0$ dynamics by retrieving the MSD, which is shown for a subset of values of $\tau_{\mathrm{p}}$ in \cref{Fig1}a. It can be seen that for long times the particles migrate diffusively ($\mathrm{MSD}\propto t$), while, for all state points, they are fastest (largest MSD) when $\tau_{\mathrm{p}}=0.005$, indicating that the departure from equilibrium occurs in a non-monotonic fashion. Such non-monotonic dependence on the persistence time 
is also consistent with previous results for quasi-hard-sphere athermal AOUPs \cite{Flenner2020,BerthierAOUP2017}. 
However, we note that for the studied state points the particles are slowest (smallest MSD) at different persistence times  (either $\tau_{\mathrm{p}}=0.0002$ or $\tau_{\mathrm{p}}=0.05$). One might be tempted to interpret this as their equivalency being lost upon departing from equilibrium, but this is in fact not the case.

\begin{figure}[ht!]
    \centering
    \vspace{0.0cm}
    \includegraphics 
    [width=9cm,height=4.5cm] 
    {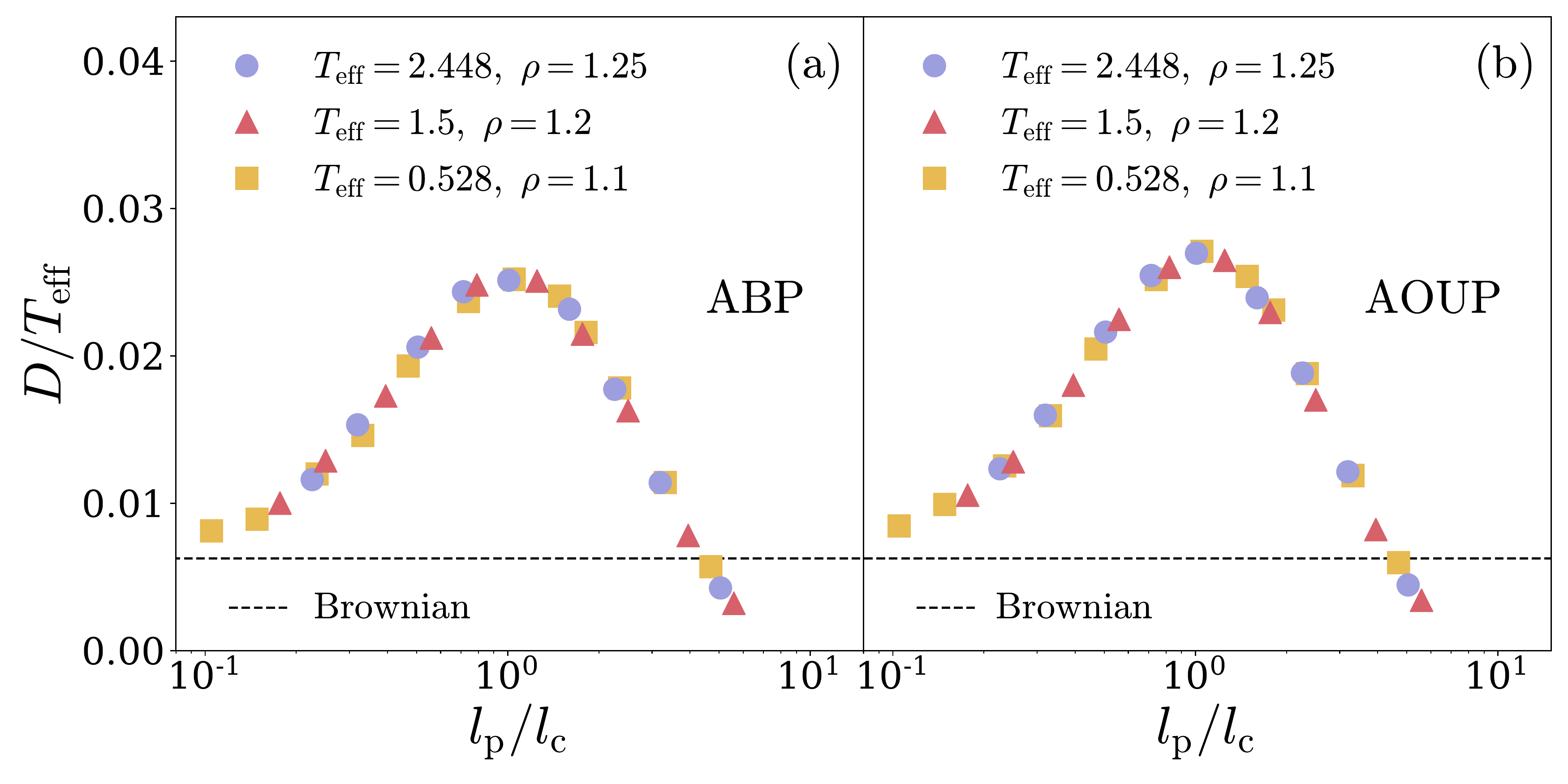} 
    \caption{The normalized self-diffusion coefficient $D/T_{\mathrm{eff}}$  
    as a function of the ratio of the persistence to the cage length $l_{\mathrm{p}}/l_{\mathrm{c}}$ for (a) athermal ABPs and (b) athermal AOUPs. 
    Both models yield almost identical results for both state points, with an optimum around the cage length $l_{\mathrm{p}}/l_{\mathrm{c}}\sim 1.0$ (with  $l_{\mathrm{c}}\sim 0.12\sigma_{\mathrm{AA}}$). The result 
    from passive Brownian dynamics simulations ($l_{\mathrm{p}}=0$) is the same for all state points, and is added as a reference (dashed line).}
    \label{Fig2}
\end{figure}

To demonstrate that there is indeed a large degree of universality hidden in the chosen state points, we explore the dependence of the dynamics on $\tau_{\mathrm{p}}$ in more detail by calculating (based on the MSDs) the self-diffusion coefficient $D=\lim_{t\to\infty} \avg{\delta r^{2}(t)}/6t$ and plotting the resulting values normalized by $T_{\mathrm{eff}}$ in \cref{Fig1}b. Interestingly, the results 
now seem almost identical except for an offset in the persistence time (explaining the differences in the MSDs). We may additionally note that for small $\tau_{\mathrm{p}}$ the self-diffusion coefficients approach, as expected, the value obtained from passive Brownian dynamics simulations (at $T=T_{\mathrm{eff}}$), while for large $\tau_{\mathrm{p}}$ it drops below this value implying that, on average, the particles migrate more slowly than their passive counterparts. 
Due to the similar shapes of the plots in \cref{Fig1}b, we anticipate that a different control parameter might be able to correct for the observed offset. Fortunately, the ABP model system comes naturally equipped with a length scale, namely the persistence length $l_{\mathrm{p}}=f\tau_{\mathrm{p}}$, and indeed when we plot the results as a function of $l_{\mathrm{p}}/l_{\mathrm{c}}$ they fully collapse (see \cref{Fig2}a). Inspection of \cref{Fig2}a then shows that the optimum value of $D$ coincides with a value $l_{\mathrm{p}}\sim 0.12\sigma_{\mathrm{AA}}$, which is entirely consistent with the size of the \textit{cage length} $l_{\mathrm{c}}$ (estimated via a nearest-neighbor analysis at a density $\rho=1.2$, see SI for details; note that  
the Lindemann rule yields $0.13$ times the particle diameter for monodisperse hard spheres \cite{Hansen2013simple}).
Since we are at relatively high densities and therefore the cage length is approximately the same for $\rho=1.1,1.2,1.25$ (see SI), this also explains why the results collapse almost perfectly. Moreover, we have verified that this collapse is robust when changing to a different $\Gamma$ value deeper in the supercooled regime (see fig.~S3a). We may also qualitatively rationalize the central role of the cage length by picturing particles trying to escape from their cage formed by neighboring  particles. This process should proceed most effectively when particles can scan all the edges for an opening as fast as possible, which occurs when the persistence length is of the same order as the cage length. In contrast, when $l_{\mathrm{p}}\ll l_{\mathrm{c}}$ it would take much longer to reach the edges of the cage, while for $l_{\mathrm{p}}\gg l_{\mathrm{c}}$ a particle tends to stick in one edge of the cage for a relatively long time.

Next, we test whether the observed behavior persists for the athermal AOUP model. For the AOUPs one can invoke the equivalency of the MSDs (see \cref{MSDsingle}) to define a persistence length as $l_{\mathrm{p}}=(3D_{f}\tau_{\mathrm{p}})^{1/2}\tau_{\mathrm{p}}$, where $(3D_{f}\tau_{\mathrm{p}})^{1/2}$ may be interpreted as the approximate average self-propulsion force. Using $l_{\mathrm{p}}/l_{\mathrm{c}}$ as our control parameter, we have plotted the calculated values of $D/T_{\mathrm{eff}}$ (for the same state points as used for the ABPs) in \cref{Fig2}b. Remarkably, the results once more collapse and the differences with the ABP model appear to be only marginal. This suggests that the specific microscopic details of these active self-propulsion mechanisms are of lesser importance in high-density systems. Intuitively, given that the particle motion becomes more impeded by repulsion at high densities and the system is starting to approach an arrested state, one would also expect the precise single-particle dynamics to become less relevant.

Finally, we mention that in previous work involving softer interaction potentials the initial increase in $D$ vanishes and only the drop after passing a critical value of $\tau_{\mathrm{p}}$ remains \cite{Flenner2020}. We have verified that when we change the power in our interaction potential from $36$ to either $18$ or $12$, the initial increase in $D$ is indeed strongly suppressed (see fig.~S3b), but the subsequent drop still occurs at approximately the same persistence length (albeit slightly smaller for increasing softness, which we believe is due to the longer range of a softer powerlaw potential). A followup study along these lines is planned for future work. 

\noindent \textit{Thermal Active Particles --} To establish whether the observed behavior undergoes qualitative changes when thermal motion is added to the active-particle models, we now compare the following three state points $[T_{\mathrm{eff}}=3.0,\ T=0.0,\ \rho=1.2]$, $[T_{\mathrm{eff}}=3.0,\ T=1.5,\ \rho=1.2]$, $[T_{\mathrm{eff}}=3.0,\ T=2.0,\ \rho=1.2]$. Note that all points have an equal effective temperature, but in one case only active motion adds to this value, while in the other two both active and passive motion contribute. 

\begin{figure}[ht!]
    \centering
    \includegraphics [width=0.38\textwidth] {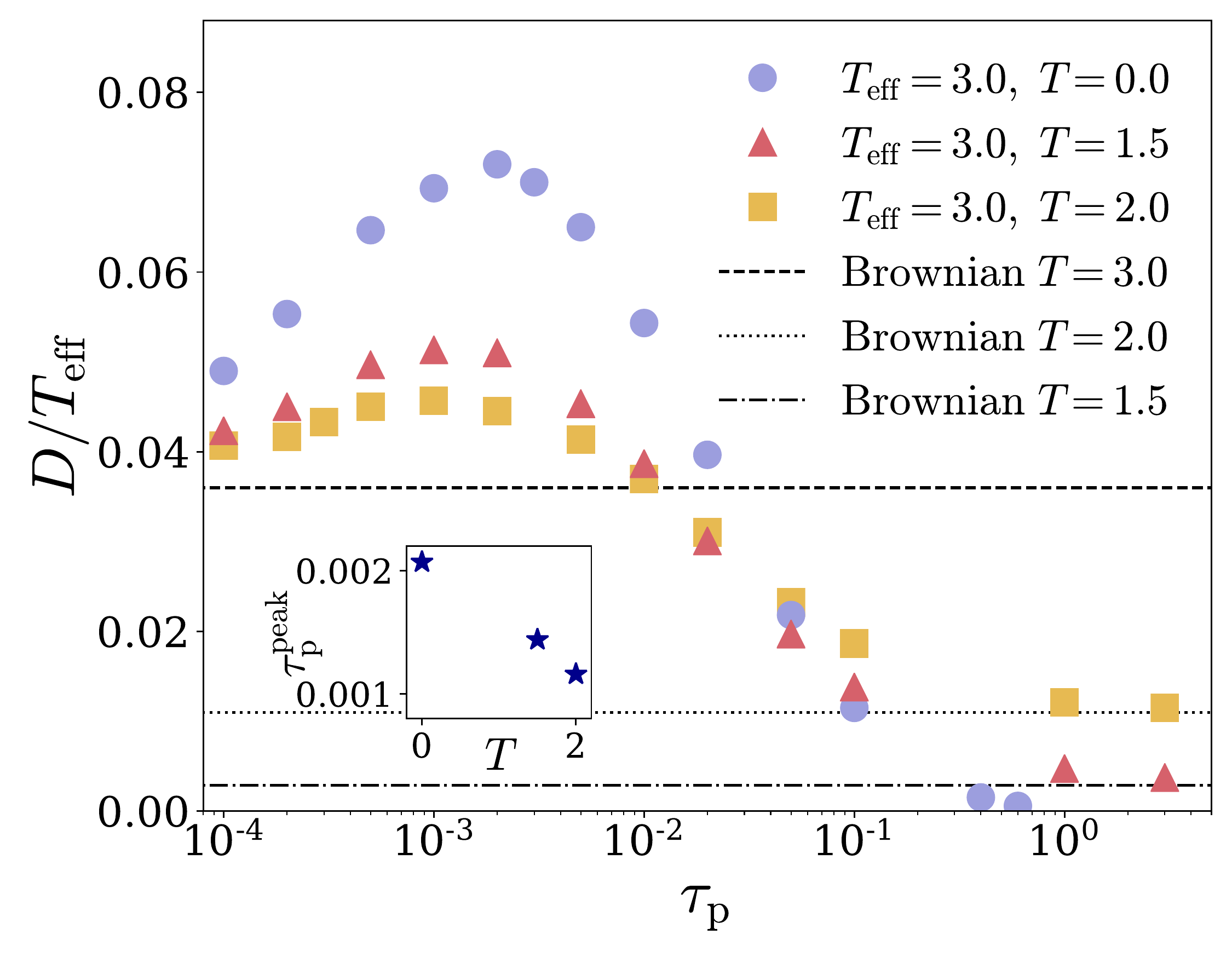} 
    \caption{The normalized self-diffusion coefficient 
    $D/T_{\mathrm{eff}}$ as a function of $\tau_{\mathrm{p}}$ for thermal ($T=1.5,2.0$) and athermal ($T=0.0$) ABPs at fixed values of $T_{\mathrm{eff}}=3.0$, $\rho=1.2$. Increasing $\tau_{\mathrm{p}}$ initially yields faster, but eventually slower, dynamics than Brownian particles at $T=3.0$ (dashed line); in the limit of large $\tau_{\mathrm{p}}$, thermal ABPs approach the $T=1.5,2.0$ passive Brownian limit 
    (dashed-dotted, dotted lines) while athermal ABPs yield $D/T_{\mathrm{eff}}\rightarrow 0$. The inset shows the persistence time $\tau_{\mathrm{p}}^{\mathrm{peak}}$ corresponding to the peak value of $D$, which decreases as a function of $T$. 
    }
    \label{Fig3}
\end{figure}

Again starting with the ABP model, we have calculated the MSDs for the mentioned state points; the corresponding values of $D/T_{\mathrm{eff}}$ are plotted as a function of $\tau_{\mathrm{p}}$ in \cref{Fig3}. Upon first glance, the qualitative shape of the thermal ABP curves look similar to its athermal counterpart. In particular, all results approach the anticipated value of Brownian particles at $T=3.0$ for small $\tau_{\mathrm{p}}$ and show the same non-monotonic behavior. 

There are, however, also some notable differences. In the limit of large $\tau_{\mathrm{p}}$, for instance, the thermal results seem to approach the same dynamics as a Brownian particle at either $T=1.5$ or $T=2.0$ (which suggests that superimposing active onto passive motion always enhances the dynamics), whereas the self-diffusion coefficient of the athermal particles manifestly goes to zero. We can explain these observations by noting that for a fixed $T_{\mathrm{a}}$, taking the limit of very large $\tau_{\mathrm{p}}$ also implies that the average self-propulsion force becomes very small. As a result, the athermal particles take increasingly long to break out of their cages, resulting in progressively slow dynamics, while the motion of the thermal particles becomes completely dominated by the passive contribution.     

Another difference is the smaller height of the peak value for the thermal ABPs, which is simply due to the active motion contributing not as much to $T_{\mathrm{eff}}$ in comparison to the athermal ABPs. More interesting is the location of the peak values. For athermal systems, we  
find that a smaller active (or effective) temperature $T_{\mathrm{a}}$ will result in an optimum self-diffusion at a larger persistence time (see \cref{Fig1}b). One might therefore expect that the optimum value in our considered thermal systems, which have a smaller value of $T_{\mathrm{a}}=1.0,1.5$ compared to its athermal analogue (with $T_{\mathrm{a}}=3.0$), would also be at a larger value for $\tau_{\mathrm{p}}$. Surprisingly, \cref{Fig3} shows the opposite. This suggests that taking $l_{\mathrm{p}}/l_{\mathrm{c}}$ as a control parameter will not result in the peak value being at the same location.

\begin{figure}[ht!]
    \centering
    \vspace{0.3cm}
    \includegraphics [width=9cm,height=4.5cm] {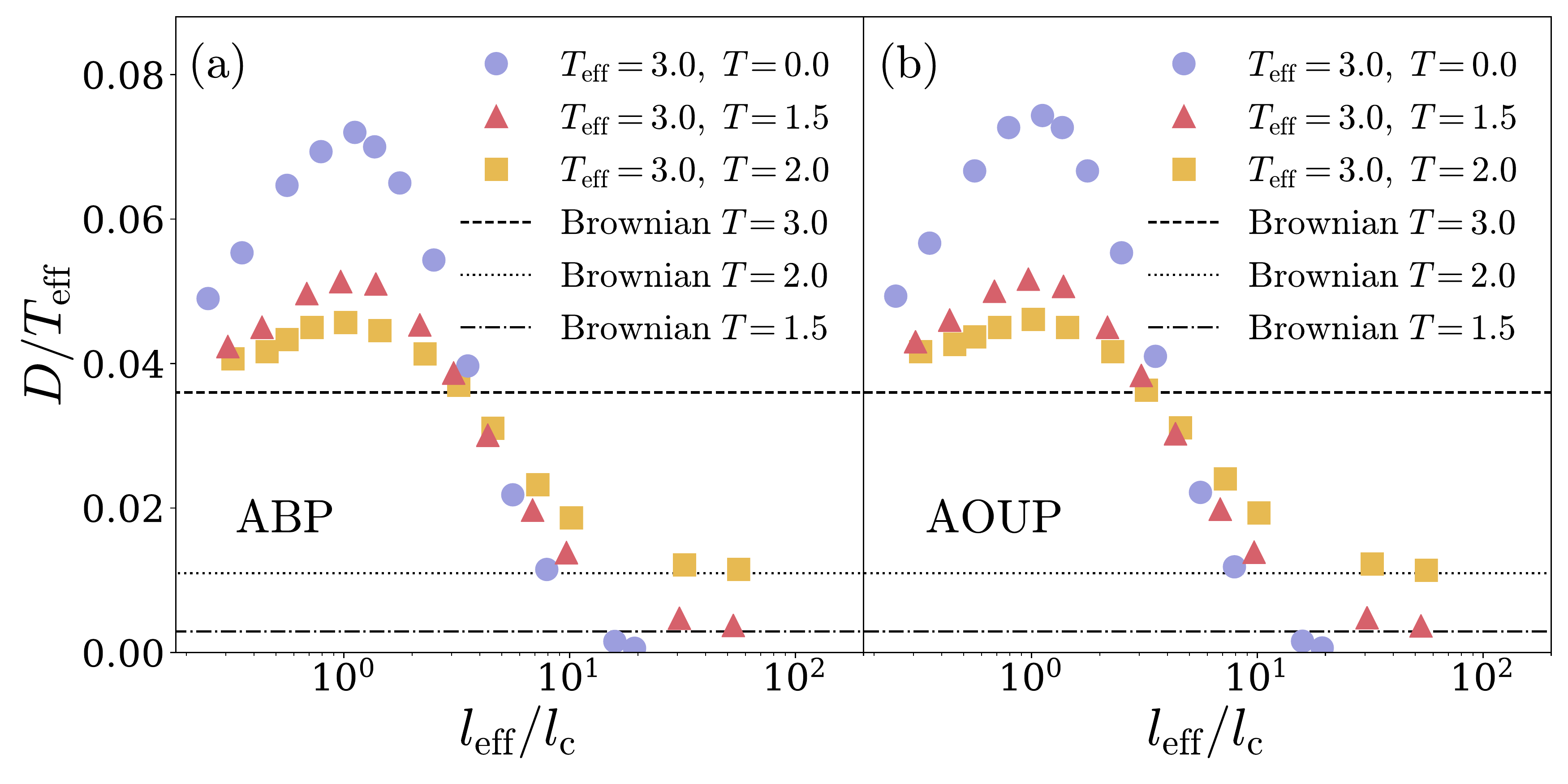} 
    \caption{The normalized self-diffusion coefficient $D/T_{\mathrm{eff}}$ as a function of the normalized effective short-time length scale $l_{\mathrm{eff}}/l_{\mathrm{c}}$ for (a) the ABP and (b) the AOUP model, at fixed values $T_{\mathrm{eff}}=3.0$, $\rho=1.2$.
    Both the thermal $T=1.5,2.0$ and athermal $T=0.0$ results have an optimum value around the cage length $l_{\mathrm{eff}}/l_{\mathrm{c}}\sim 1.0$. The passive Brownian reference at $T=1.5$ (dashed-dotted line), $T=2.0$ (dotted line), and $T=3.0$ (dashed line) is shown for comparison.}
    \label{Fig4}
\end{figure}

To resolve this discrepancy, we realize that thermal active systems are in fact governed by an additional source of motion that is inherently absent in athermal systems. Explicitly, for athermal active systems only the self-propulsion contributes to the motion of the particles and thus the short-time length scale is the persistence length. In contrast, for thermal systems, the added thermal motion enhances the length scale at short times. 
For $t<\tau_{\mathrm{p}}$ we may expand the single-particle MSD up to second order to give $\avg{\delta r^{2}(t)} \approx 6Tt + 3T_{\mathrm{a}}t^{2} / \tau_{\mathrm{p}}$. We can then introduce $l_{\mathrm{eff}}=\left(\avg{\delta r^{2}(\tau_{\mathrm{p}})}\right)^{1/2}$ (using the second-order MSD) as an enhanced effective (short-time) length scale and use it as our control parameter for thermal systems (note that for $T=0$, we still have $l_{\mathrm{eff}}=l_{\mathrm{p}}$). Indeed, when we plot the values of $D/T_{\mathrm{eff}}$ as a function of $l_{\mathrm{eff}}/l_{\mathrm{c}}$ (see \cref{Fig4}a), we find that not only do the optima coincide, but the point at which they do is again fully consistent with the cage length, i.e.\ $l_{\mathrm{eff}}/l_{\mathrm{c}}\sim1.0$. It thus seems that the addition of thermal motion changes the relevant short-time length scale, but it does not alter the physical picture of particles exhibiting the strongest enhanced dynamics when they can explore their cage as effectively as possible.

We finalize our results by mentioning that the observed behavior is once more unaltered when we interchange the thermal ABP for the thermal AOUP model. In particular, the differences in the obtained values for $D/T_{\mathrm{eff}}$ and their dependence on $l_{\mathrm{eff}}/l_{\mathrm{c}}$ are only minute (see \cref{Fig4}).

\noindent \textit{Conclusion --} To summarize, our work demonstrates that the cage length plays a vital role in the context of high-density active glassy materials. In particular, its relation to the relevant short-time active length scale, i.e.\ the sole persistence length for athermal systems or a combination of the persistence length and a diffusive length scale for thermal systems, fully determines whether the relaxation dynamics is enhanced or suppressed with respect to a Brownian system at an equal effective temperature. Indeed, an inspection of several previous findings \cite{FlennerAOUP2016,SzamelAOUP2015,BerthierAOUP2017,Flenner2020}, which have reported different departures from equilibrium, shows that their seemingly contrasting findings can be fully reconciled by identifying whether the studied parameter regimes corresponded to short-time length scales on either side or around the cage length.
Moreover, our results are robust to the microscopic details of the self-propulsion, rendering the ratio of the short-time active length scale to the cage length the crucial control parameter in both ABPs and AOUPs. We also observe that, consistent with previous work on hard-sphere ABPs \cite{DijkstraABP2013,Voigtmann2017active}, superimposing active onto thermal motion always speeds up the relaxation dynamics. To further establish the importance of this ratio, it will be interesting to study its role in more detail for strict hard-sphere systems, whose passive dynamics should be independent of temperature. In comparison, our preliminary simulations with softer interaction potentials show the same the physical picture as sketched in this work, although the initial enhancement of the dynamics becomes more suppressed. The question how this picture extends to more complex and attractive interaction potentials, as well as biologically relevant active glasses such as confluent cell models \cite{Angelini2011cell,Bi2016cell,Janssen2019active, Ruscher2020}, should be investigated in future work to ultimately fully elucidate the rich non-equilibrium glassy dynamics of active matter. 

\section*{Acknowledgments}
\noindent We acknowledge the Dutch Research Council (NWO) for financial support through a START-UP grant (V.E.D.\ and L.M.C.J.).


\bibliographystyle{apsrev4-1}
\bibliography{all}







\end{document}